\pdfoutput=1

\documentclass[11pt]{article}

\usepackage[final]{acl}

\usepackage{times}
\usepackage{latexsym}

\usepackage[T1]{fontenc}

\usepackage[utf8]{inputenc}

\usepackage{microtype}

\usepackage{inconsolata}

\usepackage{graphicx}

\usepackage{makecell}
\usepackage{algorithmicx}
\usepackage{algorithm}
\usepackage{algpseudocode}
\usepackage{times}
\usepackage{latexsym}
\usepackage{xcolor}
\usepackage{soul}
\usepackage{microtype}
\usepackage{inconsolata}
\usepackage{graphicx}
\usepackage{enumitem}
\usepackage{array}
\usepackage{url}
\usepackage{booktabs}
\usepackage{amssymb}
\usepackage{multirow}
\usepackage{pifont}
\usepackage{courier}
\usepackage{lipsum}
\usepackage{cleveref}
\usepackage{catchfile}
\usepackage[defaultcolor=magenta]{changes}
\usepackage{multicol}
\usepackage{array}
\usepackage{booktabs} 
\usepackage{catchfile} 
\usepackage{lipsum}
\usepackage{listings}
\usepackage{xcolor}
\usepackage{listingsutf8}
\usepackage{lipsum}
\usepackage{tcolorbox}

\usepackage[T1]{fontenc}  
\usepackage[utf8]{inputenc}
\usepackage{kotex}

\title{A Large-Scale Real-World Evaluation of an \\ LLM-Based Virtual Teaching Assistant}

\author{Sunjun Kweon, Sooyohn Nam, Hyunseung Lim, Hwajung Hong, Edward Choi \\
  KAIST \\
    \texttt{\{sean0042, edwardchoi\}@kaist.ac.kr}\\
  }

\begin{document}
\maketitle
\begin{abstract}
Virtual Teaching Assistants (VTAs) powered by Large Language Models (LLMs) have the potential to enhance student learning by providing instant feedback and facilitating multi-turn interactions.
However, empirical studies on their effectiveness and acceptance in real-world classrooms are limited, leaving their practical impact uncertain.
In this study, we develop an LLM-based VTA and deploy it in an introductory
AI programming course with 477 graduate students.
To assess how student perceptions of the VTA’s performance evolve over time, we conduct three rounds of comprehensive surveys at different stages of the course.
Additionally, we analyze 3,869 student–VTA interaction pairs to identify common question types and engagement patterns.
We then compare these interactions with traditional student-human instructor interactions to evaluate the VTA’s role in the learning process.
Through a large-scale empirical study and interaction analysis, we assess the feasibility of deploying VTAs in real-world classrooms and identify key challenges for broader adoption.
Finally, we release the source code of our VTA system, fostering future advancements in AI-driven education: 
\url{https://github.com/sean0042/VTA}
\end{abstract}

\section{Introduction}

Providing continuous feedback and support beyond regular class hours is essential for effective education \cite{chickering1987seven, ahea2016value}.
To address this need, educational institutions commonly rely on online learning management systems (\textit{e.g.,} Blackboard), direct email communication, or third-party discussion platforms (\textit{e.g.,} Piazza) to facilitate student-instructor interactions.
However, these tools struggle to scale in large introductory courses, where students require deeper conceptual understanding.
Effective learning in such courses depends on frequent, personalized interactions with instructors, but resource constraints make this difficult.
Instructors and TAs are often overwhelmed by the sheer volume of student inquiries, making it challenging to provide timely, personalized feedback.
Furthermore, students often hesitate to ask questions due to fear of judgment or uncertainty about whether their inquiries are appropriate \citep{ruihua2025understanding}.
This reluctance further limits access to personalized feedback and hinders conceptual learning.

The emergence of Large Language Models presents promising solution to these challenges.
LLM-based Virtual Teaching Assistants (VTAs) have shown potential to complement, and in some cases partially substitute, human instructors by providing automated responses to student inquiries \citep{hicke2023chata, wang2023chated, taneja2024jill, ahmed2024potentiality, liu2024step, kakar2024jill}.
These systems can deliver instant, contextually relevant responses and support multi-turn dialogues that foster deeper engagement.
Moreover, VTAs may help create a more inclusive learning environment by lowering barriers for students who might hesitate to ask questions in person.
Despite these potential benefits, effectiveness and acceptance of VTAs in real-world classrooms remain largely unexplored, limiting broader adoption.

In this study, we develop and deploy an LLM-based VTA in a real-world classroom at a graduate-level, introductory AI programming course in South Korea, where 477 students are enrolled.
To assess students’ perceived effectiveness and usefulness of the VTA, we conduct three rounds of surveys—pre-deployment, mid-deployment, and post-deployment—tracking how their perceptions evolve over time.
These surveys evaluate the VTA’s perceived helpfulness, trustworthiness, response appropriateness, and comfort level compared to a human instructor.
Additionally, we collect and analyze 3,869 question-response interactions between students and the VTA, identifying engagement patterns and comparing them with traditional student-human interactions.
By integrating survey insights with interaction analysis, this study offers a comprehensive evaluation of VTAs in real-world classrooms, highlighting their potential to enhance student learning while addressing challenges for broader implementation.

\section{Related Works}

The development of VTAs for answering student inquiries has gained significant attention in recent years.
One of the pioneering efforts, \citet{goel2018jill}, introduced a VTA leveraging IBM’s Watson APIs to classify student questions and retrieve relevant answers from episodic memory.
However, its inability to generate contextually adaptive responses limited its utility \citep{eicher2018jill}.
Recent advances in LLMs have enhanced VTA capabilities.
Studies such as \citet{hicke2023chata}, \citet{wang2023chated}, and \citet{ahmed2024potentiality} demonstrate the effectiveness of LLM-based VTAs in various educational settings.
Notable real-world deployments include JeepyTA at the University of Pennsylvania \citep{liu2024step} and Jill Watson at Georgia Tech \citep{kakar2024jill}, illustrating the potential of VTAs in classrooms.
These systems typically use GPT-based models \citep{brown2020language} and leverage retrieval-augmented generation \citep{lewis2020retrieval} to ensure contextually relevant responses aligned with course content.
Our study builds upon this prior research while addressing several key limitations of earlier works:

\paragraph{Limited Large-Scale Evaluations:}
Many existing studies evaluate VTAs using LLM evaluations or small-scale surveys, offering limited empirical validation.
Our study addresses this gap through large-scale surveys with 477 students, enabling a comprehensive assessment of perceived helpfulness, trustworthiness, response appropriateness, and comfort level—metrics selected with reference to \citet{han2023recipe}—compared to a human instructor across three survey rounds.
Furthermore, our study spans an entire semester, allowing a longitudinal perspective on student perceptions over time.
\paragraph{Lack of Interaction-Level Analysis:} 
Most prior research focuses on high-level evaluations, rarely analyzing the actual interactions between students and VTAs.
We conduct an in-depth analysis of 3,869 student-VTA interactions, identifying engagement patterns and comparing them to traditional student-human interactions.
\paragraph{Limited Accessibility and Reproducibility:}
Many existing VTA systems are not publicly available, limiting their adoption despite demonstrated efficacy.
To facilitate broader accessibility and customization, we publicly release the source code of our VTA system, providing a practical resource for future research and educational applications.

\section{Deployment Background}

In the Fall semester of 2024, we deployed an LLM-based VTA in an introductory AI programming course at a graduate school in South Korea. 
The deployment lasted for 14 weeks, from September to December.
The course integrated machine learning and artificial intelligence theories with hands-on programming in PyTorch.
Live online sessions were held twice a week: one for theory lectures and another for coding exercises, both conducted in English.
Students were required to complete three major programming projects to strengthen their theory understanding and implementation skills.
The instructional team consisted of one professor responsible for theory lectures and course management, supported by eight TAs who facilitated coding sessions and project guidance.
Course materials—including lecture slides (PDFs) and coding resources (Jupyter Notebooks)—were shared via the school’s online Blackboard system before each class.
Sessions were recorded for later review, and important announcements were posted on Blackboard.
While critical or grade-related questions were addressed during live sessions or via Blackboard’s Q\&A section, students were encouraged to use the VTA for general inquiries related to course content and coding assistance.

The course enrolled 477 students from 30 different departments. 
Students' academic levels spanned doctoral (20.6\%), master's (78.9\%), and undergraduate (0.5\%) programs. 
The class also included international students from 22 countries (see Appendix \ref{appendix:demographics} for details).
To evaluate the VTA’s impact, we conducted three mandatory survey rounds—before, during, and after deployment (see Appendix \ref{appendix::survey_questions} for the survey questions).
While survey participation was required for course completion, students were assured that their responses would not affect their grades, ensuring honest feedback.
Of the 477 students, 472 consented to participate under Institutional Review Board (IRB) approval, allowing us to analyze their survey responses and student-VTA interaction logs.

\section{VTA Architecture}

    \begin{figure}[t]
    \includegraphics[width=0.48\textwidth,trim={184 123 342 112}, clip]{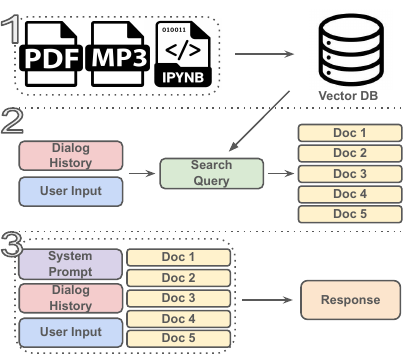}
    \caption{Overview of the VTA architecture. (1) The system processes educational materials into a vector database, (2) retrieves relevant documents based on students' queries, and (3) generates responses.}
    \label{fig1}
    \end{figure}

The VTA developed for this study was implemented using three open-source Python libraries: LangChain, Streamlit, and LangSmith.
LangChain serves as the core framework for building the LLM-based chatbot for the VTA, enabling Retrieval-Augmented Generation \citep{lewis2020retrieval} from a vector database constructed using processed course materials.
Streamlit provides the web interface and LangSmith is used for storing and analyzing conversation histories between the students and the VTA.
The overall architecture of the VTA is illustrated in Figure \ref{fig1}.
The system operates based on the following key components:

\paragraph{1. Building and Updating the Vector Database}

The VTA relies on three main types of reference materials for RAG: theory lecture PDFs (\textit{.pdf}), practice code files (\textit{.ipynb}), and lecture recodings (\textit{*.mp3}).
The audio part of the lecture recordings were transcribed into text using OpenAI's \texttt{Whisper-1} model \citep{radford2023robust}.
To ensure efficient search during the retrieval phase, long documents were segmented into 2,048-token chunks, with a 256-token overlap between chunks to maintain contextual continuity.
Each chunk was prefixed with the lecture date and title to provide additional context.
Vector embeddings for these chunks were then generated using OpenAI's \texttt{text-embedding-3-large} model \citep{neelakantan2022text}.
The resulting embeddings were stored in a Faiss-based vector database \citep{johnson2019billion,douze2024faiss}, allowing for fast similarity computation during document retrieval.
The vector database was updated after each class session.
Over the course of the semester, 59 lecture materials—including PDFs, Jupyter Notebooks, and class recordings—were collected, resulting in 1,502 chunks stored in the database.

\paragraph{2. Retrieving Documents using Search Query} 

To perform RAG, the VTA first embeds the user’s query and retrieves the most relevant documents from the vector database.
However, embedding only the latest question may not always capture the full conversational context, especially in multi-turn dialogues.
For example, if a student first asks, \textit{`When is Project 1 due?'} and later follows up with, \textit{'What is the task about?'} simply embedding the second question might fail to retrieve relevant documents since \textit{`Project 1'} was only mentioned in the previous turn.
To address this, VTA first generates a context-aware search query before retrieval.
Specifically, the \texttt{gpt-4o-mini} model processes the dialog history along with the latest question to produce a consolidated query—for instance, \textit{`Project 1 task contents'}. 
The full prompt used for query generation is provided in Appendix Figure \ref{prompt:search_query}.

Once generated, the search query is embedded using the same OpenAI model (\texttt{text-embedding-3-large}) and compared with stored document embeddings to retrieve the most relevant materials.
A key hyperparameter in this process is the number of retrieved documents (\texttt{k}).
While retrieving more documents can improve accuracy, it also increases computational cost and latency.
After empirical evaluations, we found that retrieving the top five documents provides the best trade-off for our use case.

\paragraph{3. Retrieval Augmented Response Generation}

Once the top five relevant documents are retrieved, the VTA generates a response using the \texttt{gpt-4o-mini} model.
The model takes as input the system prompt, the dialog history, the student’s latest question, and the retrieved documents to generate a contextually informed answer.
The system prompt includes essential class logistics along with the current date and time, obtained via Python’s datetime module.
This ensures responses to time-sensitive queries, such as \textit{`What is the answer for the quiz we did in last week's practice?'}. 
The full prompt details are provided in Appendix Figure \ref{prompt:response_generation}.

\paragraph{4. Serving VTA and Storing Dialog History}

The VTA is deployed via a Streamlit web interface, allowing students to access it through a shared link.
To ensure secure access, students must enter their student ID, which is verified against stored credentials managed through Streamlit’s secret key feature.
A screenshot of the VTA interface is provided in Appendix \ref{appendix:interface}.
All conversation logs are recorded using LangSmith for analysis.
Each log entry includes the student ID, conversation history, submitted queries, VTA-generated responses, timestamps, and details of the retrieved documents.

\section{VTA Usage Analysis}

\subsection{Usage Overview}

    \begin{table}[h]
    \centering
    \setlength{\heavyrulewidth}{1.0pt}
    \resizebox{1.0\columnwidth}{!}{
        \newcommand{\mytable}{\begin{tabular}{c|c|c|c}
\toprule 
Group & Usage Range & \# of Users & Total Q\&A Count \\ \midrule
A & ≥ 100 times & 6 & 1,154 \\
B & 18 ≤ times \textless 100 & 53 & 1,872 \\
C & 5 ≤ times \textless 18 & 69 & 604 \\
D & \textless 5 times & 107 & 239 \\
E & No usage & 237 & - \\
\midrule
Total & - & 472 & 3,869 \\
\bottomrule
\end{tabular}}
            \begin{tabular}{lc}
\hline
\textbf{Reported Limitation} & \multicolumn{1}{l}{\textbf{Count}} \\ \hline
Hallucination or incorrect answers & 10 \\
Slow response time & 22 \\
Failure to follow instructions precisely & 11 \\
Difficulty retrieving course-related content & 8 \\
No issues reported & 69 \\
Others & 10 \\ \hline
\end{tabular}
    }
    \caption{
Categorization of students based on their usage frequency with the VTA. 
    }
    \label{table1}
    \end{table}

The VTA was deployed over a 14-week lecture period with an operational cost of approximately \$180, covering API usage and conversation log storage.
Among 472 students, nearly 50\% engaged with the VTA at least once, resulting in 916 conversations and 3,869 individual interactions (Q\&A exchanges).
Student interaction volumes varied significantly, ranging from a single query to a maximum of 375.
To analyze usage patterns, students were grouped into five categories based on interaction frequency, as summarized in Table \ref{table1}.
Quartile-based thresholds were used: Q2 (median) at 5 interactions and Q3 at 18.
Q1 was observed at 2 interactions, but its small gap from single-use cases led to its exclusion as a separate category.
Students with over 100 interactions were classified as outliers.
The following analysis examines engagement trends and behaviors across these groups.

\subsection{Impact of Academic Background and Prior Knowledge on Usage}

To better understand which students engaged most actively with the VTA, we analyzed usage patterns based on academic background and prior knowledge, specifically coding experience and machine learning knowledge familiarity.
For academic background, students were classified into two groups: \textit{Computer Science-Related} and \textit{Non-Computer Science-Related} disciplines.
Students from non-computer science fields showed significantly higher engagement, with 80\% of high-frequency users (Groups A and B) in this category.

    \begin{table}[h]
    \centering
    \setlength{\heavyrulewidth}{1.0pt}
    \resizebox{1.0\columnwidth}{!}{
        \newcommand{\mytable}{
\begin{tabular}{c|c|c|c|c}
\toprule
- & None & Beginner & Intermediate & Advanced \\
\midrule
Coding Experience & 62.2 & 11.2 & 5.5 & 4.5 \\
ML Knowledge & 23.6 & 11.1 & 7.1 & 3.0 \\
\bottomrule
\end{tabular}
}
            \begin{tabular}{lc}
\hline
\textbf{Reported Limitation} & \multicolumn{1}{l}{\textbf{Count}} \\ \hline
Hallucination or incorrect answers & 10 \\
Slow response time & 22 \\
Failure to follow instructions precisely & 11 \\
Difficulty retrieving course-related content & 8 \\
No issues reported & 69 \\
Others & 10 \\ \hline
\end{tabular}
    }
    \caption{
Average VTA interactions by prior coding experience and Machine Learning knowledge
    }
    \label{table2}
    \end{table}


In addition, the pre-deployment survey asked about students’ prior experience in coding and machine learning, categorizing them into four levels: None, Beginner, Intermediate, and Advanced.
As summarized in Table \ref{table2}, students with no prior coding experience showed the highest engagement with the VTA, averaging 62.2 interactions, followed by beginners (11.2), intermediates (5.5), and advanced users (4.5).
A similar pattern appeared regarding prior machine learning knowledge, with students lacking experience utilizing the VTA most frequently.
These findings suggest the VTA served as a valuable learning aid, particularly for students needing additional support.

\subsection{Comparison with Student-Instructor Engagement}

    \begin{table}[h]
    \centering
    \setlength{\heavyrulewidth}{1.0pt}
    \resizebox{1.0\columnwidth}{!}{
        \newcommand{\mytable}{\begin{tabular}{c|c|c}
\toprule 
Question Type & Human TA (Last Year) & Virtual TA (This Year) \\
\midrule
Coding Practice& 9.0\% & 10.4\% \\
ML Theories & 8.3\% & 35.0\% \\
Projects & 66.4\% & 39.7\% \\
Course Operation & 15.3\% & 9.7\% \\
\bottomrule
\end{tabular}}
            \begin{tabular}{lc}
\hline
\textbf{Reported Limitation} & \multicolumn{1}{l}{\textbf{Count}} \\ \hline
Hallucination or incorrect answers & 10 \\
Slow response time & 22 \\
Failure to follow instructions precisely & 11 \\
Difficulty retrieving course-related content & 8 \\
No issues reported & 69 \\
Others & 10 \\ \hline
\end{tabular}
    }
    \caption{
Distribution of student inquiries across four categories for both VTA and human instructor interactions.
    }
    \label{table3}
    \end{table}

    \begin{table*}[t]
    \centering
    \setlength{\heavyrulewidth}{1.0pt}
    \resizebox{2.0\columnwidth}{!}{
        \newcommand{\mytable}{\begin{tabular}{c|cccc|cccc|cccc|ccc}
\toprule
 & \multicolumn{4}{c|}{Helpfulness} & \multicolumn{4}{c|}{Trustworthiness} & \multicolumn{4}{c|}{Appropriateness} & \multicolumn{3}{c}{Comfortableness} \\ \midrule
Group & Pre & Mid & Post & Human & Pre & Mid & Post & Human & Pre & Mid & Post & Human & Pre & Mid & Post \\ \midrule
All & 3.64 & 3.60 & 3.54 & 3.96 & 3.27 & 3.44 & 3.51 & 4.38 & 3.71 & 3.80 & 3.92 & 4.07 & 0.58 & 0.58 & 0.65 \\
A & 3.50 & 3.62 & 3.66 & 3.66 & 3.50 & 3.52 & 3.50 & 4.33 & 4.00 & 4.02 & 3.83 & 3.67 & 0.83 & 0.77 & 0.83 \\
B & 3.58 & 3.72 & 3.76 & 4.04 & 3.31 & 3.39 & 3.53 & 4.47 & 3.61 & 3.78 & 3.98 & 4.16 & 0.55 & 0.68 & 0.71 \\
C & 3.56 & 3.71 & 3.77 & 3.77 & 3.27 & 3.56 & 3.62 & 4.32 & 3.74 & 3.95 & 4.05 & 3.95 & 0.62 & 0.68 & 0.73 \\
D & 3.72 & 3.55 & 3.26 & 4.06 & 3.23 & 3.12 & 3.42 & 4.38 & 3.73 & 3.73 & 3.81 & 4.13 & 0.56 & 0.62 & 0.56 \\ \bottomrule
\end{tabular}}
            \begin{tabular}{lc}
\hline
\textbf{Reported Limitation} & \multicolumn{1}{l}{\textbf{Count}} \\ \hline
Hallucination or incorrect answers & 10 \\
Slow response time & 22 \\
Failure to follow instructions precisely & 11 \\
Difficulty retrieving course-related content & 8 \\
No issues reported & 69 \\
Others & 10 \\ \hline
\end{tabular}
    }
    \caption{
Survey Results on Students' Perceptions of the VTA Across Deployment Phases and Comparison with Human Instructors.
    }
    \label{table4}
    \end{table*}

Analyzing how students interacted with VTA versus human instructors can offer valuable insights into its role in learning.
We examined 3,869 student–VTA Q\&A exchanges from this year and 144 student–instructor interactions from the same course last year, which used a third-party Q\&A platform.
The stark contrast in volume—students asked over 25 times more questions to VTA—suggests that it provided a more approachable and accessible way to seek help.
We categorized all questions into four types: coding, theory, project-related, and course administration (see Table \ref{table3}).
While project-related queries were the most common in both cases, theory-related questions were notably more frequent with the VTA.
This suggests that students may have felt more comfortable engaging in deeper conceptual discussions with the VTA, likely due to its on-demand availability and non-judgmental nature (see Section \ref{survey_analysis}).

In addition to the content of interactions, the nature of student engagement plays a crucial role in shaping the learning experience.
To explore whether students felt a sense of connection with the VTA similar to that with human instructors, we analyzed social interactions characterized by interpersonal exchanges and rapport—such as casual greetings, expressions of gratitude, humor, and anthropomorphic remarks.
Each conversation was processed using a large language model to automatically identify these relational elements.
Of the 916 recorded conversations, 123 (13\%) included such social cues, while the remaining 793 (87\%) were purely informational.
Students who engaged in relational dialogue interacted with the VTA an average of 27.8 times, compared to just 11.4 times among those who did not.
These findings suggest that students who sought to establish a friendly and comfortable atmosphere with the VTA—mirroring human-like interaction—tended to engage with it more frequently.
Future work could explore how such dynamics influence student engagement and motivation in AI-assisted learning.

\section{Survey Analysis}
\label{survey_analysis}

Understanding how students perceive the VTA is crucial for evaluating its effectiveness in real-world classrooms.
To this end, we conducted three rounds of surveys—before deployment (pre), during deployment (mid), and after deployment (post)—to track changes in student perceptions over time.
The survey assessed four key dimensions:
\begin{itemize}[leftmargin=3.5mm]
    \item \textbf{Helpfulness} : How useful students found the VTA’s responses (1 = Not helpful, to 5 = Very helpful). \vspace{-2mm}
    \item \textbf{Trustworthiness} : The degree to which students trusted the VTA’s answers (1 = Do not trust at all, to 5 = Fully trust). \vspace{-2mm}
    \item \textbf{Appropriateness} : How well the VTA’s response style (e.g., tone, clarity) aligned with students' expectations (1 = Very inappropriate, to 5 = Very appropriate). \vspace{-2mm}
    \item \textbf{Comfortableness} : How comfortable students felt asking questions to the VTA compared to human TAs (-1 = Less comfortable, 0 = Same, +1 = More comfortable).
\end{itemize}

For the first three aspects, students also rated their experiences with human instructors to establish a comparative baseline. 
The survey results, summarized in Table \ref{table4}, reveal how student perceptions evolved over time and how the VTA compared to human instructors in key evaluation metrics. 
Overall, student evaluations of the VTA improved from pre-deployment to post-deployment except for Helpfulness from Group D. 
Below, we provide a detailed analysis of each metric.

\paragraph{Helpfulness}

The overall perception of the VTA’s helpfulness showed a slight decline from pre-deployment (3.64) to mid-deployment (3.60) and post-deployment (3.54). 
However, among high-frequency users (Groups A, B, and C), there was a statistically significant improvement in the Helpfulness score after sustained usage (\textit{p} = 0.043). 
This suggests that extended interaction enhances students’ recognition of the VTA’s usefulness.
In contrast, Group D exhibited a decline in Helpfulness ratings after use (Pre: 3.72~$\rightarrow$~Post: 3.26), which may indicate that these students initially had higher expectations that were not fully met. Notably, Group D also rated human TAs the highest in helpfulness (4.06) among all groups, suggesting that they placed greater value on the support provided by human instructors. 
As a result, they may have initially expected a similar level of support from the VTA but found it lacking after limited use (2.2 times on average), leading to a decline in their perceived helpfulness.

\paragraph{Trustworthiness}

The perceived trustworthiness of the VTA’s responses increased after deployment, suggesting that while students were initially skeptical, they gradually found its answers to be more accurate and consistent than expected. 
However, trust in the VTA remained lower compared to human instructors, indicating that students still viewed human instructors as more reliable. 
This underscores a key limitation of VTAs—while they can still provide useful and contextually relevant information, they have yet to match the perceived dependability of human instructors in educational settings.

\paragraph{Appropriateness}

Student evaluations of the VTA’s appropriateness—assessing factors such as tone, clarity, and response structure—showed a positive trend throughout the deployment.  
Unlike other metrics, appropriateness received relatively high ratings from the pre-deployment stage, indicating that students generally expected the VTA’s response style acceptable.
Notably, appropriateness was the metric with the smallest gap between post-deployment VTA ratings and human instructor ratings, suggesting that students found the VTA’s response style relatively comparable to that of human instructors.

\paragraph{Comfortableness}
To assess how comfortable students felt interacting with the VTA compared to human TAs, we analyzed their responses before and after deployment (with scores closer to -1 indicating a preference for human TAs, 0 indicating no preference, and scores closer to 1 indicating a preference for the VTA).  
Before deployment, the average comfort score across all students was 0.58, suggesting that a significant number of students initially expected the VTA to be more comfortable to interact with than human instructors.  
While the overall comfort score increased slightly from pre- to post-deployment, the change was not statistically significant (\textit{p} = 0.097).  
However, among high-frequency users (Groups A, B, and C), a significant increase in comfort was observed (\textit{p} = 0.000748), indicating that frequent users became progressively more at ease using the VTA over time.

    \begin{table}[h]
    \centering
    \setlength{\heavyrulewidth}{1.0pt}
    \resizebox{1.0\columnwidth}{!}{
        \newcommand{\mytable}{\begin{tabular}{c|c|c|c}
\toprule 
- & Comfortable (Pre) & Comfortable (Post) & Avg Usage \\
\midrule
Refrain? (Yes) & 0.69 & 0.76 & 13.2 \\
Refrain? (No) & 0.42 & 0.47 & 7.8 \\
\bottomrule
\end{tabular}}
            \begin{tabular}{lc}
\hline
\textbf{Reported Limitation} & \multicolumn{1}{l}{\textbf{Count}} \\ \hline
Hallucination or incorrect answers & 10 \\
Slow response time & 22 \\
Failure to follow instructions precisely & 11 \\
Difficulty retrieving course-related content & 8 \\
No issues reported & 69 \\
Others & 10 \\ \hline
\end{tabular}
    }
    \caption{
Comfort scores and VTA usage based on prior hesitation to ask human instructors.
    }
    \label{table5}
    \end{table}

Additionally, a notable insight emerged from our pre-survey question:  
\textit{“Have you ever refrained from asking a question to a human instructor due to discomfort, fear of burdening them, or concern that your question might seem silly?”}.
58\% of students responded “Yes” (had refrained), while 42\% responded “No” (had not refrained). 
Table \ref{table5} presents the average comfort scores and VTA usage for these two groups. 
A key observation is that students who had previously refrained from asking human instructors reported higher comfort scores both pre- and post-deployment (Pre: 0.69~$\rightarrow$~Post: 0.76) compared to those who had not refrained (Pre: 0.42~$\rightarrow$~Post: 0.47). 
This suggests that students who were initially hesitant to engage with human instructors found the VTA a more comfortable alternative.  
Furthermore, usage patterns aligned with this trend—students who had refrained from asking human instructors exhibited a higher average VTA usage (13.2 interactions) compared to those who had not refrained (7.8 interactions).  
These findings highlight the potential of VTAs in reducing psychological barriers to asking questions, particularly for students who might otherwise hesitate to engage with human instructors.

\section{Limitations}

To further investigate the limitations of the VTA in educational settings, we included the following question in the survey: “\textit{Did you encounter any issues or limitations while using the VTA?}”
To ensure the feedback reflected meaningful engagement, we limited our analysis to students whose number of interactions with the VTA met or exceeded the median usage threshold (five interactions).
Students with fewer than five interactions were excluded, as their limited exposure was deemed insufficient to reliably assess the system’s limitations.
Respondents could select from six options: four predefined issues—(1) hallucinated or incorrect answers, (2) slow response time, (3) failure to follow instructions, and (4) difficulty retrieving course-related content—alongside a “no issues” option and an open-ended “other” category.
Multiple selections were allowed. 
Table~\ref{table6} summarizes the distribution of reported issues.

    \begin{table}[h]
    \centering
    \setlength{\heavyrulewidth}{1.0pt}
    \resizebox{1.0\columnwidth}{!}{
        \newcommand{\mytable}{\begin{tabular}{lc}
\hline
\textbf{Reported Limitation} & \multicolumn{1}{l}{\textbf{Count}} \\ \hline
Hallucination or incorrect answers & 10 \\
Slow response time & 22 \\
Failure to follow instructions precisely & 11 \\
Difficulty retrieving course-related content & 8 \\
No issues reported & 69 \\
Others & 10 \\ \hline
\end{tabular}}
            \begin{tabular}{lc}
\hline
\textbf{Reported Limitation} & \multicolumn{1}{l}{\textbf{Count}} \\ \hline
Hallucination or incorrect answers & 10 \\
Slow response time & 22 \\
Failure to follow instructions precisely & 11 \\
Difficulty retrieving course-related content & 8 \\
No issues reported & 69 \\
Others & 10 \\ \hline
\end{tabular}
    }
    \caption{
Summary of reported issues among students with frequent VTA usage.
    }
    \label{table6}
    \end{table}

A substantial proportion of students selected the “no issues” option, suggesting that many encountered no problems during their interactions with the VTA.
Among those who did report issues, the most common concern was slow response time.
However, empirical comparisons with public LLMs such as ChatGPT revealed no significant difference in output generation latency for equivalent prompts.
We attribute this perception to the VTA’s lack of output streaming. Unlike standard LLM interfaces, which display partial responses as they are generated, the VTA delivers the complete output at once.
This likely led students accustomed to streaming interfaces to perceive the system as slower.
Incorporating streaming functionality could address this concern.

Other reported issues—such as failures to follow instructions and hallucinated or incorrect responses—were less frequent but align with known limitations of current LLMs.
Given the modular design of the VTA, improvements in the underlying LLM architecture can be readily adopted to enhance instruction-following and factual accuracy.
A smaller number of students reported difficulties in retrieving course-relevant content. These cases often involved content that was commonly discussed in class, indicating potential weaknesses in the retrieval mechanism. The current implementation uses dense vector similarity for retrieval. To improve recall and precision, future versions of the VTA could adopt hybrid retrieval strategies (e.g., combining dense vectors with sparse models like BM25) or expand the document candidate pool to improve coverage.

    Finally, open-ended responses in the “other” category surfaced system-level and presentation-related issues. Examples included formatting problems such as rendering errors in markdown equations and repeated words across lines. These were not observed during internal testing and likely stem from implementation bugs that can be addressed through routine debugging. Additionally, some students noted that VTA responses felt overly constrained to course materials and lacked broader explanatory context. This limitation may be alleviated by adjusting the system prompt to encourage more comprehensive and context-aware answers.

\section{Conclusion}

We developed and deployed an LLM-based Virtual Teaching Assistant in a graduate-level AI programming course with 472 students, evaluating its impact through large-scale surveys and analysis of 3,869 student interactions. 
Results showed that students' perceptions of the VTA improved across multiple dimensions—helpfulness, trustworthiness, appropriateness, and comfort—with the most notable gains among frequent users and those hesitant to approach human instructors. 
The VTA not only supported scalable, personalized assistance but also contributed to a more inclusive learning environment.
However, the VTA did not fully match the perceived reliability or depth of support provided by human instructors, highlighting current limitations in LLM-based educational tools. 
Moreover, since our deployment focused on a programming-oriented course, its effectiveness in other domains with different cognitive demands remains to be tested. To support future research, we publicly release the source code of our VTA system.

\section*{Ethics Statement}

The study was approved by the Institutional Review Board of KAIST (Approval Number: KH2024-276) and adhered to ethical guidelines for research involving human subjects.

\section*{Acknowledgments}
This work was supported by the KAIST Center for Exellence in Learning \& Teaching, the Institute of Information \& Communications Technology Planning \& Evaluation (IITP) grant (No.RS-2019-II190075, No.RS-2024-00338140, No.RS-2025-02304967) and National Research Foundation of Korea (NRF) grant (NRF-2020H1D3A2A03100945), funded by the Korea government (MSIT).


\bibliography{custom}

\newpage

\appendix

\onecolumn

\section{Prompts}
\label{appendix:prompts}

    \begin{figure}[H]
    \centering
    \scalebox{0.90}{
    \input{prompt/search_query}
    }
    \caption{Prompt Template for Search Query Generation}
    \label{prompt:search_query}
    \end{figure}

    \begin{figure}[H]
    \centering
    \scalebox{0.90}{
    \input{prompt/response_generate}
    }
    \caption{Prompt Template for VTA Response Generation}
    \label{prompt:response_generation}
    \end{figure}

\newpage

\section{Student Statistics}
\label{appendix:demographics}

\begin{figure}[!h]   
    \centering
    \includegraphics[width=0.8\textwidth,trim={11 3 10 10}, clip]{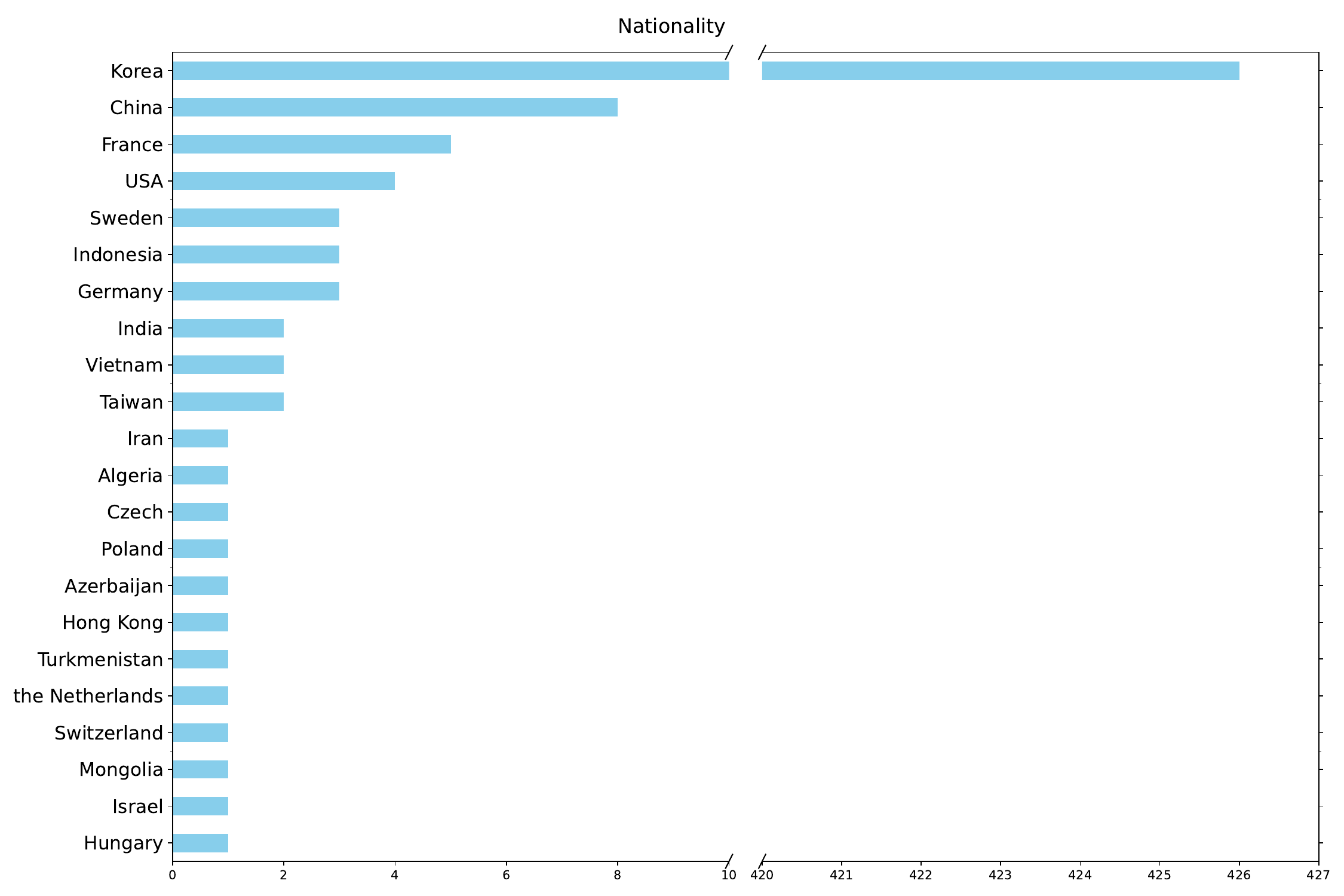}
    \caption{Student Statistics : Nationality}
    \label{fig2}
    \end{figure}

\begin{figure}[!h]   
    \centering\includegraphics[width=0.86\textwidth,trim={10 0 5 0}, clip]{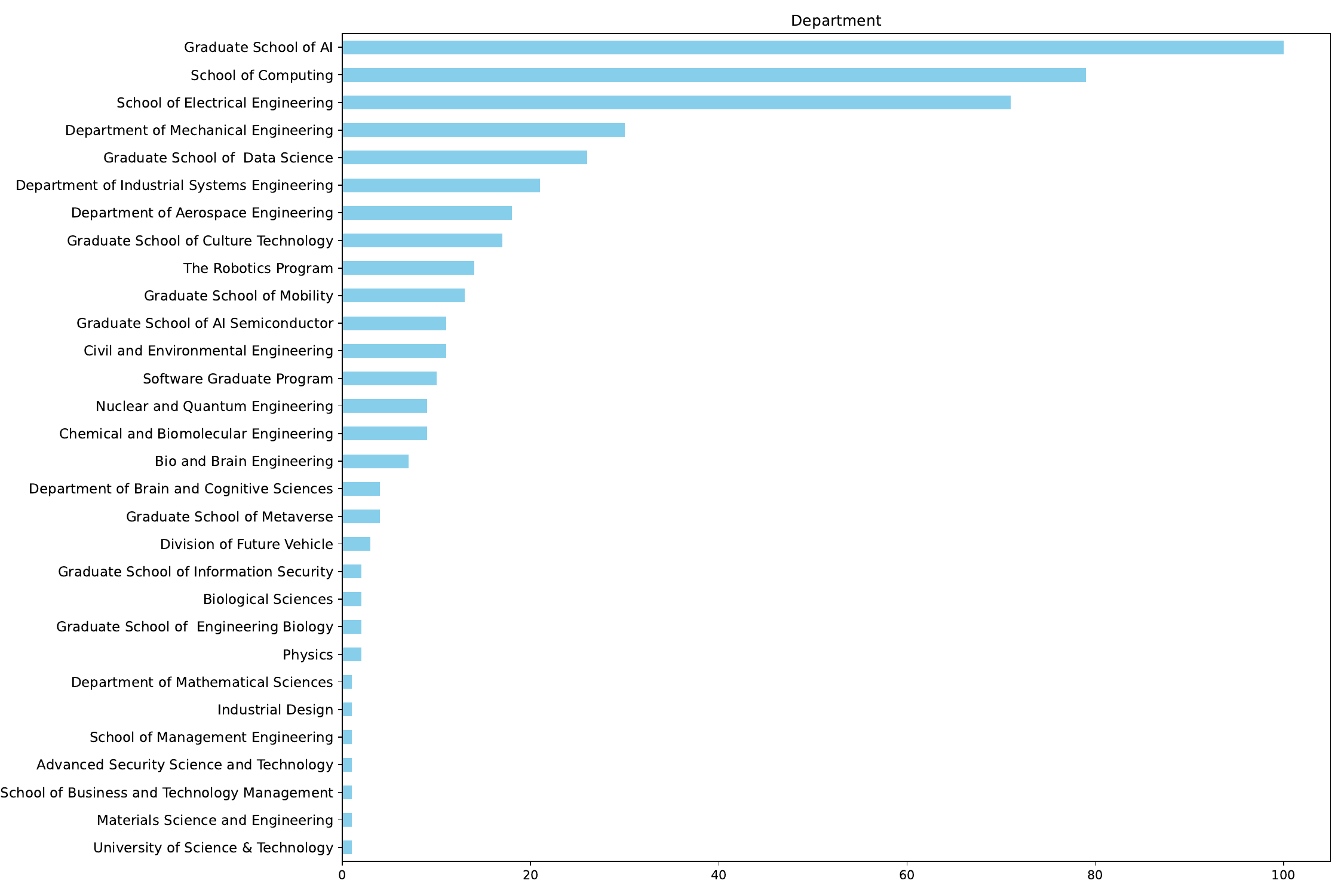}
    \caption{Student Statistics : Departments}
    \label{fig4}
    \end{figure}

Figures \ref{fig2} and \ref{fig4} present the demographic distribution of the 472 students enrolled in the course.
Figure \ref{fig2} illustrates the students' nationalities, showing that they come from 22 different countries. The majority of students are from Korea, followed by China, France, and the United States.
Figure \ref{fig4} displays the distribution of students across various academic departments. The largest groups belong to the Graduate School of AI, School of Computing, and School of Electrical Engineering, with students also coming from diverse fields such as mechanical engineering, aerospace engineering, and industrial design.

\newpage

\section{VTA Interface Screenshot}
\label{appendix:interface}

\begin{figure}[h]   
    \centering
    \includegraphics[width=0.9\textwidth,trim={180 5 180 50}, clip]{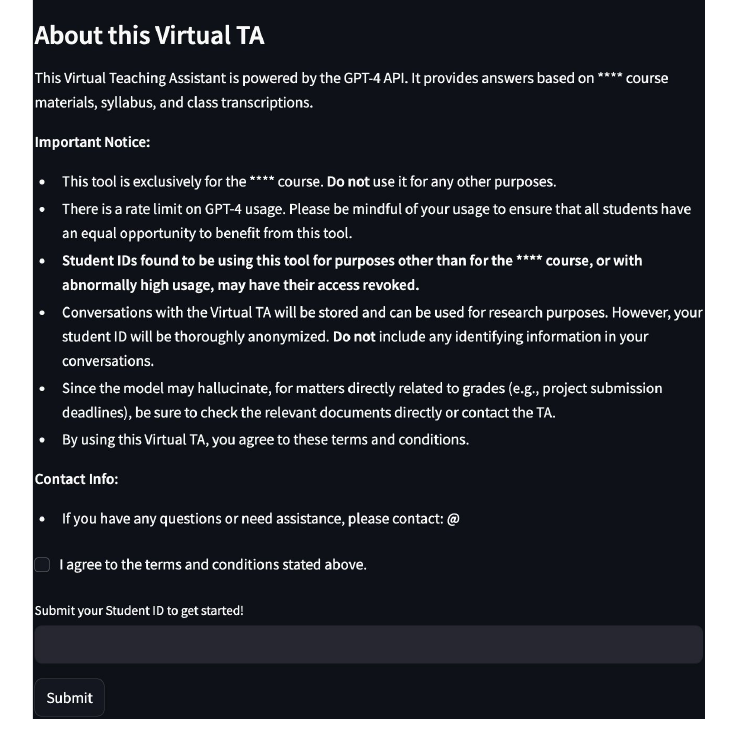}
    \caption{Initial VTA screen with a usage guide}
    \label{fig5}
    \end{figure}

\begin{figure}[!h]   
    \centering
    \includegraphics[width=0.9\textwidth,trim={240 10 240 10}, clip]{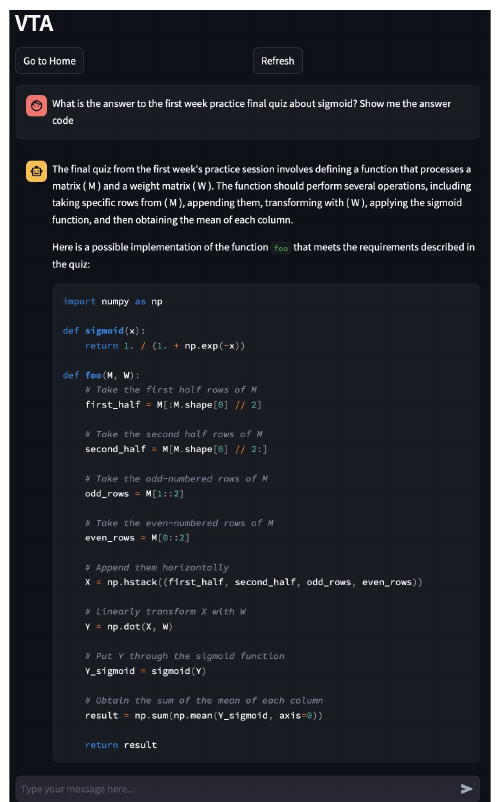}
    \caption{VTA Chatbot interface displayed after student ID verification.}
    \label{fig6}
    \end{figure}

Figures \ref{fig5} and \ref{fig6} show screenshots of the VTA deployed in this study. 
Figure \ref{fig5} displays the initial screen that appears when accessing the VTA via the shared link, providing a brief usage guide. 
After entering their student ID, users gain access to the chatbot interface, shown in Figure \ref{fig6}, which includes example questions and responses.

\newpage

\section{Survey Questions}
\label{appendix::survey_questions}
\subsection{Pre-deployment Survey}
\begin{enumerate}
    \item \textbf{What is your current academic status?}  
        \begin{itemize}
            \item Undergraduate
            \item Master’s Student
            \item PhD Student
        \end{itemize} 
    \item \textbf{Prior Coding Experience}  
        \begin{itemize}
            \item None: I have never written any code
            \item Beginner: I have taken at least one course in any programming language (e.g. C++, Java, Python)
            \item Intermediate: I have taken (or knowledgeable in) Data Structure and Algorithms courses.
            \item Advanced: I have done projects in advanced courses such as Compiler, Operating Systems, Embedded Systems or Networks.
        \end{itemize}
    \item \textbf{Prior Machine Learning Knowledge}  
        \begin{itemize}
            \item None: I don’t have any experience/knowledge in machine learning
            \item Beginner: I am familiar with basic data analysis such as regression, classification or clustering
            \item Intermediate: I have taken (or knowledgeable in) at least one undergrad-level machine learning course
            \item Advanced: I have taken (or knowledgeable in) advance deep learning courses such as Stanford’s CS231n (Computer Vision) and CS224n (Natural Language Processing)
        \end{itemize}    
    \item \textbf{Have you ever refrained from asking a question to a human instructor due to discomfort, fear of burdening them, or concern that your question might seem silly?}  
        \begin{itemize}
            \item Yes
            \item No
        \end{itemize}  

    \item \textbf{How helpful do you expect the responses from an LLM-based TA to be?}  
        \begin{itemize}
            \item Not helpful at all (1)
            \item Slightly helpful (2)
            \item Moderately helpful (3)
            \item Helpful (4)
            \item Very helpful (5)
        \end{itemize}  

    \item \textbf{How much would you trust the responses from an LLM-based TA?}  
        \begin{itemize}
            \item Do not trust at all (1)
            \item Slightly trust (2)
            \item Moderately trust (3)
            \item Trust (4)
            \item Fully trust (5)
        \end{itemize}  

    \item \textbf{How appropriate do you expect the style of the responses (clarity, tone, etc.)?}  
        \begin{itemize}
            \item Very inappropriate (1)
            \item Slightly inappropriate (2)
            \item Moderately appropriate (3)
            \item Appropriate (4)
            \item Very appropriate (5)
        \end{itemize}  

    \item \textbf{Compared to a human TA, how comfortable would you be asking questions to an LLM-based TA?}  
        \begin{itemize}
            \item More uncomfortable (-1)
            \item About the same (0)
            \item More comfortable (1)
        \end{itemize}  
\end{enumerate}

\subsection{Mid-deployment Survey}

\begin{enumerate}
    \item \textbf{In the first survey, you responded to “How helpful do you expect the responses from an LLM-based TA to be?” After using it, what is your opinion on above question?}  
        \begin{itemize}
            \item Not helpful at all (1)
            \item Slightly helpful (2)
            \item Moderately helpful (3)
            \item Helpful (4)
            \item Very helpful (5)
        \end{itemize} 

    \item \textbf{In the first survey, you responded to “How much would you trust the responses from an LLM-based TA?” After using it, what is your opinion on above question?}  
        \begin{itemize}
            \item Do not trust at all (1)
            \item Slightly trust (2)
            \item Moderately trust (3)
            \item Trust (4)
            \item Fully trust (5)
        \end{itemize}  

    \item \textbf{In the first survey, you responded to “How appropriate do you expect the style of the responses (clarity, tone, etc.)?” After using it, what is your opinion on above question?}  
        \begin{itemize}
            \item Very inappropriate (1)
            \item Slightly inappropriate (2)
            \item Moderately appropriate (3)
            \item Appropriate (4)
            \item Very appropriate (5)
        \end{itemize}  

    \item \textbf{In the first survey, you responded to “Compared to a human TA, how comfortable would you be asking questions to an LLM-based TA?” After using it, what is your opinion on above question?}  
        \begin{itemize}
            \item More uncomfortable (-1)
            \item About the same (0)
            \item More comfortable (1)
        \end{itemize}

\end{enumerate}

\subsection{Post-deployment Survey}

\begin{enumerate}
    \item \textbf{After using LLM-TA, what is your final opinion on the question "How helpful do you find the responses from an LLM-TA"?}  
        \begin{itemize}
            \item Not helpful at all (1)
            \item Slightly helpful (2)
            \item Moderately helpful (3)
            \item Helpful (4)
            \item Very helpful (5)
        \end{itemize} 

    \item \textbf{After using LLM-TA, what is your final opinion on the question "How much did you trust the responses from an LLM-based TA?"?}  
        \begin{itemize}
            \item Do not trust at all (1)
            \item Slightly trust (2)
            \item Moderately trust (3)
            \item Trust (4)
            \item Fully trust (5)
        \end{itemize}  

    \item \textbf{After using LLM-TA, what is your final opinion on the question "How appropriate did you find the style of the responses (clarity, tone, etc.) to be?"?}  
        \begin{itemize}
            \item Very inappropriate (1)
            \item Slightly inappropriate (2)
            \item Moderately appropriate (3)
            \item Appropriate (4)
            \item Very appropriate (5)
        \end{itemize}  

    \item \textbf{After using LLL-TA, what is your final opinion on the question "Compared to a human TA, how comfortable did you find asking questions to an LLM TA?""?}  
        \begin{itemize}
            \item More uncomfortable (-1)
            \item About the same (0)
            \item More comfortable (1)
        \end{itemize}  

    \item \textbf{How much would you recommend the LLM-TA to prospective students of this class?}  
        \begin{itemize}
            \item Not at all recommend
            \item Slightly recommend
            \item Moderately recommend
            \item Highly recommend
            \item Strongly recommend
        \end{itemize}  

    \item \textbf{Compared to general purpose LLMs (e.g. chatGPT, Claude), do you agree that the LLA-TA is more specialized for this course?}  
        \begin{itemize}
            \item Strongly Disagree
            \item Disagree
            \item Neutral
            \item Agree
            \item Strongly Agree
        \end{itemize}  

\end{enumerate}

\end{document}